**Z. Jouini, O.O. Kurakevych, H. Moutaabbid and Y. Le Godec** (Paris, France)
**M. Mezouar** (Grenoble, France)
**N. Guignot** (Gif-sur-Yvette, France)

## Phase boundary between Na-Si clathrates of structures I and II at high pressures and high temperatures.

*Understanding the covalent clathrate formation is a crucial point for the design of new superhard materials with intrinsic coupling of superhardness and metallic conductivity. Silicon clathrates have the archetype structures that can serve an existant model compounds for superhard clathrate frameworks "Si-B", "Si-C", "B-C" and "C" with intercalated atoms (e.g. alkali metals or even halogens) that can assure the metalic properties. Here we report the in situ and ex situ studies of high-pressure formation and stability of clathrates $Na_8Si_{46}$ (structure I) and $Na_{24+x}Si_{136}$ (structure II). Experiments have been performed using standard Paris-Edinburgh cells (opposite anvils) up to 6 GPa and 1500 K. We have established that chemical interactions in Na-Si system and transition between two structures of clathrates occur at temperatures below silicon melting. The strong sensitivity of crystallization products to the sodium concentration have been observed. A tentative diagram of clathrate transformations has been proposed. At least up to ~6 GPa, $Na_{24+x}Si_{136}$ (structure II) is stable at lower temperatures as compared to $Na_8Si_{46}$ (structure I).*

The techniques of high pressure and high temperature (HPHT) appear to be the tools of choice for the synthesis of novel superhard [1,2] and other useful materials [3] with optimized properties. Although such synthesis usually leads to the significant decrease of the reaction volume, (1) the strong changes of crystal chemistry [4,5], chemical reactivity [6-8], crystallization and growth rate [9], chemical thermodynamics [10-13] and thermophysical properties [14,15]; and (2) the possibility of direct interactions of components [7], direct phase transformations [16] and nanostructuring [17] allow considering this method as promising for future materials science [18-21], outside traditional limit of diamond and boron nitride materials.

The modern design of advanced superhard materials is often based on the phases with crystal structures typical for the elements of second period (diamond, boron and borides) [1,2,4,5,16,17]. At the same time, theory predicts the superhard frameworks "B-C" and "C" that have the crystal structures more typical for larger analogues, for example for silicon compounds (e.g. clathrates, zeolites). With intercalated atoms (e.g. alkali metals or halogenes) one can expect not only high hardness - intrinsic for covalent nets of rigid bonds - but also metallic conductivity. These materials are highly desired for industry in order to replace the sintered powders of superhard (diamond, cBN) and conductive (Co) materials. Silicon compounds with clathrate [22,23] and zeolite [23,24] structure are archetypical and can serve as an existant point of departure.

Silicon clathrates with structures often isotypical to gas hydrates of type I ($A_8Si_{46}$) and II ($A_{24}Si_{136}$) [22] have the most rigid covalent framework as compared to other clathrates (with frameworks produced by other elements of group 14, silicon oxide or water), since carbon clathrates are not known at present time. So they are the hardest known clathrates with highest elastic moduli. In particular, the hardness, just like the bulk modulus and other elastic constants, is close to that of conventional diamond silicon Si-I, even in the case of empty-cage $Si_{136}$. The rigidity of silicon framework is defined by Si-Si bondings, and both intercalated metals and voids does not influence it much. Known silicon clathrates are thermoelectric and superconducting materials [19]. The compounds of alkali and alkaline earth metals can be obtained by thermal decomposition of Zintl type silicides (e.g. $Na_4Si_4$, $Ba_2Si_4$). However, the stoichiometric bulks of best properties can be obtained only at the *p-T* domain(s) of thermodynamic stability of clathrates, i.e. at HPHT conditions [23], where such clathrates can be formed directly from the elements. However, the HPHT phase transformations in clathrate-forming systems have not been studied so far. Recent discovery of orthorhombic allotrope of silicon with promising for optoelectronics quasidirect bandgap, $Si_{24}$ [24] - that forms from high-pressure $NaSi_6$ clathrate [23] - has attracted attention to the Na-Si system under pressure. However, in order to design B- and C- bearing clathrates, understanding of mutual stability of the Na-Si compounds remains a challenging problem that has been partially solved in the present letter.

Here we experimentally studied the phase boundary between sodium clathrate materials ($Na_8Si_{46}$, type I and $Na_{24+x}Si_{136}$, a high-pressure form of type II with $x$ up to 6.5) and their formation from Na+Si mixture and decomposition. The high-pressure sytheses and transformations of silicon clathrates up to 6 GPa and 1500 K were carried out using the standard Paris-Edinburgh cells (opposite-anvil geometry), previously well characterized during other *in situ p-T* studies [6,25]. The initial Na+Si mixture (commercial products, handled in Ar atmosphere) or clathrate II (HP synthesis, handled in air) were placed into a cylindrical capsules from hexagonal boron nitride (volume ~3mm³) and sealed with a BN cap. The capsules were placed within the graphite cylinders (furnaces) assuring the electro-resistive heating. Pyrophyllite or boron-epoxy gaskets were employed as a pressure trasmitting medium. Additional Teflon rings were placed around the gaskets to avoid the lost of cell pressure due to material leakage (especially in the case of pyrophyllite). Typical experiments were performed between 3 and 6 GPa at temperatures from 900K to 1500K (heating time from 0.5 to 3h) using standard 10/3.5 assembly (Fig. 1a). Pressure calibration for boron-epoxy gasket (pressure medium) has originally been performed *in situ* using 300 K equation of state (EOS) of hBN. It was confirmed in our experiments by EOS of Si and Na and can be presented as $P_{cell}(GPa) = 0.0125 \times P_{oil}(bars) - 8.85 \times 10^{-6} \times [P_{oil}(bars)]^2$ (typical pressure reproducibility ±0.5 GPa). In the case of pyrophyllite gasket the cell pressure - oil pressure curve was the same within reproducibility (according to the EOS of Si). Temperature calibration (*T*- power curve) has been obtained using a number of *in house* experiments with K-type thermocouple (hBN sample). *In situ* observations of Si melting at known pressure (Fig. 1b) allowed to refine the calibration to $T(K) = 300+3.25 \times power(W)$ (typical temperature reproducibility ±50 K). Similar values of power (±5 W) required for Si melting were observed in the case of both boron-epoxy and pyrophyllite gaskets.

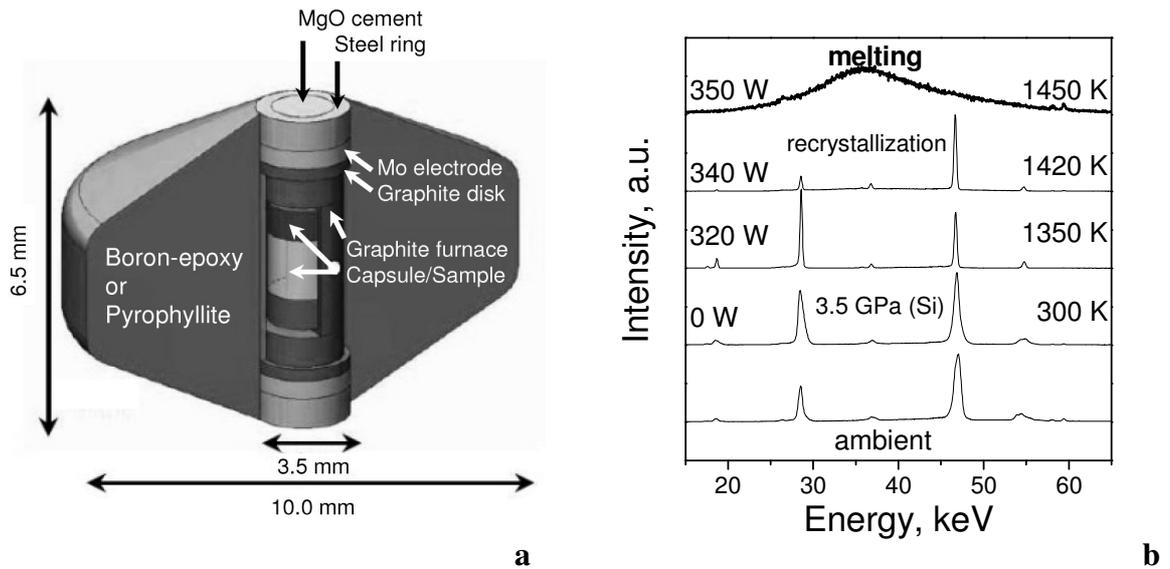

**Figure 1.** (*a*) Drawing, materials and dimensions of standard 10/3.5 assembly of high-pressure cell for opposite-anvil Paris –Edinburgh press. (*b*) *In situ* data (beamline PSICHE, synchrotron SOLEIL) on the power of Si melting at 3.5 GPa in boron-epoxy gasket (Si gauge [26]). 350 W corresponds to ~1450 K according to the melting curve of Si [27].

Our *in situ* experiments at PSICHÉ beamline of synchrotron SOLEIL have shown that silicon melts at power of ~350 W at 3.5 GPa (Fig. 1b), well in agreements with expected 1450 K according to thermocouple calibration curve and melting curve of silicon [27]. All transformations between clathrates (and reactions of Na or $Na_4Si_4$ with Si) typically occur below silicon melting curve.

*In situ* data on synthesis (ESRF, ID 27) is presented on Fig. 2a. At 4 GPa the mixture of Na+Si has been heated. First Na reacts with Si leading, probably, to disordered $Na_4Si_4$ with broad XRD reflextions. At high temperatures clathrate II crystallizes, while further temperature increase leads to clathrate I formation. This sequence of transformations is well in agreement with *ex situ* experiments described in ref. [28]. We should note that the duration of our experiments was of ~ 0.5-3 h, while in ref. [28] the corresponding powders were obtained within 20 min., and single crystals - during ~1 h. This fact is indicative of the temperature - and not time - impact on the relative stabilities of structures I and II: type II is stable at lower temperature, while type I at higher temperatures, at least at pressures of ~4-5 GPa.

The transformation diagram based on the *in situ* (Fig. 2a) and *ex situ* (Tab. 1) experiments is presented on Fig. 2b. In general, our data is well in agreement with previous results on synthesis of polycrystalline samples in Ta capsules by Kurakevych *et al*. [23], and on single-crystal growth in boron nitride capsules by Yamanaka *et al*. [28]. The different time scales of experiments allow speaking about the closeness, at least in part, of this diagram to the isopleth section (~15at% Na) of the *p-T-x* phase diagram of Na-Si system. However, strong non-stoichiometry of type II clathrate (in fact, the large cages, a characteristic feature of clathrate II, can host two atoms of sodium under pressure) seems to render the situation more complicated. The dashed circle on Fig. 2b indicates the most questioned *p-T* region at present time, where the competition of crystallizing phases is remarkable and very sensitive to the sodium concentration. At higher pressures, $NaSi_6$ clathrate become stable, at least at high temperatures and sodium concentrations close to the compound stoichiometry [23]. Below ~2 GPa, no clathrates can be formed in the system, instead, Zintl compound $Na_4Si_4$ seems to be the only stable compound [23] that can be also recovered [28].

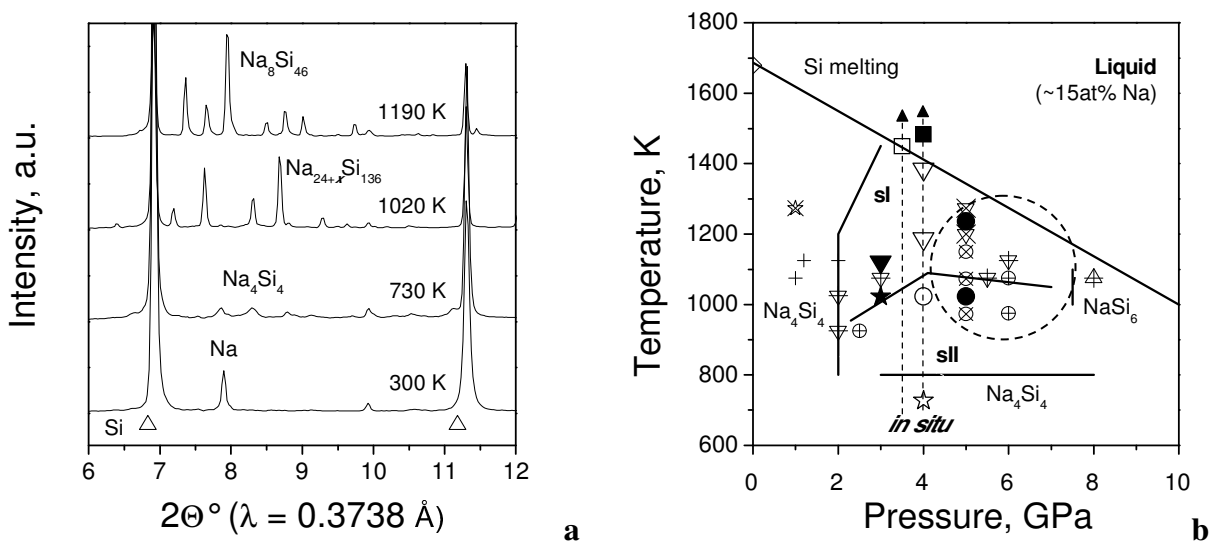

**Figure 2.** (*a*) *In situ* XRD data on phase/chemical transformations in the Na+Si mixture (~15at% of Na) during heating at 4 GPa obtained at ID27 (ESRF). (*b*) The clahtrate transformation diagram in the Na-Si system. Most of the experimental points correspond to 15-20 at% of Na. Symbols represent compounds recovered or observed *in situ* ($Na_4Si_4$ - ★, sI-$Na_8Si_{48}$ - ▼, sII-$Na_{24+x}Si_{136}$ - ●, $NaSi_6$ - ▲, and melt - ■). Solid symbols indicate the results of our recovery experiments, open symbols - our *in situ* data. For comparison, the *ex situ* data from other works are also shown: symbols crossed with + for data from Kurakevych *et al*. [23], and with ✗ from Yamanaka *et al*. [28]. Solid lines are given to guide the eyes.

Finally, a tentative diagram of formation and stability of Na-Si clathrates has been proposed. At low pressures (below 2 GPa), no clathrate synthesis is possible. Between 3 and 6 GPa one can obtain $Na_4Si_4$, clathrate II and clathrate I with temperature increase. Above 7 GPa the clathrate $NaSi_6$ become stable at high temperatures. The obtained results allow us to suggest that most probable candidates for boron- and carbon-bearing clathrates are doped type I compounds (e.g. $Na_8C_ySi_{46-y}$ or $Na_8B_xSi_{46-x}$ with $x$ up to 8 corresponding to Zintl phase with highest hardness), since the temperature of stability of clathrate I is the highest among other structural types and close to the onset of B and C atom's mobility in possible precursors that may be used as sources of boron and carbon (boron silicides, SiC, etc.).

**Table 1.** Experimental details and phase composition of recovered samples (*ex situ* data).

| No | Composition | Pressure | Temperature | Recovered phases | | | |
|---|---|---|---|---|---|---|---|
| | | | | Si | clathrate I | clathrate II | $Na_4Si_4$ |
| Z004 | Si + Na (15at%) | 5 GPa | 1230 K | + | | ++ | |
| Z007 | HP clathrate II (+NaCl) | 3 GPa | 1120 K | + | ++ | | |
| Z008 | | 3 GPa | 1020 K | + | | +(?) | ++ |
| Z009 | | 5 GPa | 1020 K | + | | ++ | +(?) |
| Z014 | Si | 4 GPa | 1480 K | melt | | | |

**Acknowledgements.** The *in situ* experiments on high-pressure synthesis and phase transformation probing were performed on the ID27 beamline at the European Synchrotron Radiation Facility (ESRF), Grenoble, France. We also acknowledge SOLEIL for provision of synchrotron radiation facilities and we would like to thank J.-P. Itié for assistance in using PSICHÉ beamline. We also thank to B. Baptiste (IMPMC) for assistance in sample characterization by powder X-ray difraction, and to V.A. Mukhanov for synthesis of starting high-pressure clathrate II in toroid apparatus at LSPM-CNRS. The PhD work of Z. Jouini was financially supported by the Ministry of Higher Education and Scientific Research of Tunisia.

**References**
1. *Kurakevych, O.O.* Superhard phases of simple substances and binary compounds of the B-C-N-O system: from diamond to the latest results (a Review) // J. Superhard Mater. - 2009. - **31**, N 3. - P. 139-157.
2. *Kurakevych, O.O., Solozhenko, V.L.* High-pressure route to superhard boron-rich solids. // High Press. Res. - 2011. - **31**, N 1. - P. 48-52.
3. *McMillan, P.F.* Chemistry of materials under extreme high pressure-high-temperature conditions. // Chem. Comm. - 2003. - **2003**, N 8. - P. 919-923.
4. *Solozhenko, V.L., Kurakevych, O.O.* Chemical interaction in the B-BN system at high pressures and temperatures. Synthesis of novel boron subnitrides. // J. Solid State Chem. - 2009. - **182**, N 6. - P. 1359-1364.
5. *Oganov, A.R., Chen, J., Gatti, C., Ma, Y., Ma, Y., Glass, C.W., Liu, Z., Yu, T., Kurakevych, O.O., Solozhenko, V.L.* Ionic high-pressure form of elemental boron. // Nature - 2009. - **457**, N 7231. - P. 863-867.
6. *Strobel, T.A., Kurakevych, O.O., Kim, D.Y., Le Godec, Y., Crichton, W., Guignard, J., Guignot, N., Cody, G.D., Oganov, A.R.* Synthesis of β-$Mg_2C_3$: A Monoclinic High-Pressure Polymorph of Magnesium Sesquicarbide. // Inorg. Chem. - 2014. - **53**, N 13. - P. 7020-7027.
7. *Kurakevych, O.O., Strobel, T.A., Kim, D.Y., Cody, G.D.* Synthesis of $Mg_2C$: a magnesium methanide. // Angew. Chem. Int. Ed. - 2013. - **52**, N 34. - P. 8930-8933.
8. *Kurakevych, O.O., Le Godec, Y., Strobel, T.A., Kim, D.Y., Crichton, W.A., Guignard, J.* High-Pressure and High-Temperature Stability of Antifluorite $Mg_2C$ by in Situ X-ray Diffraction and ab Initio Calculations. // J. Phys. Chem. C - 2014. - **118**, N 15. - P. 8128-8133.
9. *Solozhenko, V.L., Kurakevych, O.O., Sokolov, P.S., Baranov, A.N.* Kinetics of the Wurtzite-to-Rock-Salt Phase Transformation in ZnO at High Pressure. // J. Phys. Chem. A - 2011. - **115**, N 17. - P. 4354-4358.

10. *Turkevich, V.Z., Stratiichuk, D.A., Tonkoshkura, M.A., Bezhenar, N.P.* Thermodynamic calculation of the Al-B system at pressures to 8 GPa. // J. Superhard Mater. - 2014. - **36**, N 6. - P. 437-439.
11. *Turkevich, V.Z., Solozhenko, V.L.* Thermodynamic calculation of the B-C system at pressures to 24 GPa. // J. Superhard Mater. - 2014. - **36**, N 5. - P. 358-360.
12. *Solozhenko, V.L., Kurakevych, O.O., Turkevich, V.Z., Turkevich, D.V.* Phase Diagram of the B-BN System at 5 GPa. // J. Phys.Chem. B - 2010. - **114**, N 17. - P. 5819-5822.
13. *Solozhenko, V.L., Kurakevych, O.O.* Equilibrium p-T Phase Diagram of Boron: Experimental Study and Thermodynamic Analysis. // Sci. Rep. - 2013. - **3**, N n/a - art. 2351.
14. *Kurakevych, O.O., Solozhenko, V.L.* Thermoelastic equation of state of boron suboxide $B_6O$ up to 6 GPa and 2700 K: Simplified Anderson-Grüneisen model and thermodynamic consistency. // J. Superhard Mater. - 2014. - **36**, N 4. - P. 270-278.
15. *Le Godec, Y., Mezouar, M., Kurakevych, O.O., Munsch, P., Nwagwu, U., Edgar, J.H., Solozhenko, V.L.* Equation of state of single-crystal cubic boron phosphide. // J. Superhard Mater. - 2014. - **36**, N 1. - P. 61-64.
16. *Solozhenko, V.L., Kurakevych, O.O., Andrault, D., Le Godec, Y., Mezouar, M.* Ultimate metastable solubility of boron in diamond: Synthesis of superhard diamond-like $BC_5$. // Phys. Rev. Lett. - 2009. - **102**, N 6. - P. 015506.
17. *Solozhenko, V.L., Kurakevych, O.O., Le Godec, Y.* Creation of Nanostuctures by Extreme Conditions: High-Pressure Synthesis of Ultrahard Nanocrystalline Cubic Boron Nitride. // Adv. Mater. - 2012. - **24**, N 12. - P. 1540-1544.
18. *Yamanaka, S.* Silicon clathrates and carbon analogs: high pressure synthesis, structure, and superconductivity. // Dalton Trans. - 2010. - **39**, N 8. - P. 1901-1915.
19. *Yamanaka, S., Enishi, E., Fukuoka, H., Yasukawa, M.* High-pressure synthesis of a new silicon clathrate superconductor, $Ba_8Si_{46}$. // Inorg. Chem. - 2000. - **39**, N 1. - P. 56-58.
20. *Baranov, A.N., Kurakevych, O.O., Tafeenko, V.A., Sokolov, P.S., Panin, G.N., Solozhenko, V.L.* High Pressure Synthesis and Luminescent Properties of Cubic ZnO/MgO Nanocomposites. // J. Appl. Phys. - 2010. - **107**, N 7. - P. 073519.
21. *Baranov, A.N., Sokolov, P.S., Kurakevych, O.O., Tafeenko, V.A., Trots, D., Solozhenko, V.L.* Synthesis of rock-salt MeO-ZnO solid solutions (Me = $Ni^{2+}$, $Co^{2+}$, $Fe^{2+}$, $Mn^{2+}$) at high pressure and high temperature. // High Press. Res. - 2008. - **28**, N 4. - P. 515-519.
22. *Kasper, J.S., Hagenmul.P, Pouchard, M., Cros, C.* Clathrate Structure of Silicon and $Na_xSi_{136}$ (X=11). // Science - 1965. - **150**, N 3704. - P. 1713-1714.
23. *Kurakevych, O.O., Strobel, T.A., Kim, D.Y., Muramatsu, T., Struzhkin, V.V.* Na-Si Clathrates Are High-Pressure Phases: A Melt-Based Route to Control Stoichiometry and Properties. // Cryst. Grow. Des. - 2013. - **13**, N 1. - P. 303-307.
24. *Kim, D.Y., Stefanoski, S., Kurakevych, O.O., Strobel, T.A.* Synthesis of an open-framework allotrope of silicon. // Nat. Mater. - 2015. - **14**, N 2. - P. 169–173.
25. *Solozhenko, V.L., Kurakevych, O.O., Le Godec, Y., Brazhkin, V.V.* Thermodynamically Consistent p-T Phase Diagram of Boron Oxide $B_2O_3$ by in Situ Probing and Thermodynamic Analysis. // J. Phys. Chem. C - 2015. - **119**, N 35. - P. 20600-20605.
26. *Hu, J.Z., Spain, I.L.* Phases of silicon at high pressure. // Solid State Comm. - 1984. - **51**, N 5. - P. 263-266.
27. *Kubo, A., Wang, Y., Runge, C.E., Uchida, T., Kiefer, B., Nishiyama, N., Duffy, T.S.* Melting curve of silicon to 15 GPa determined by two-dimensional angle-dispersive diffraction using a Kawai-type apparatus with X-ray transparent sintered diamond anvils. // J. Phys. Chem. Solids - 2008. - **69**, N 9. - P. 2255-2260.
28. *Yamanaka, S., Komatsu, M., Tanaka, M., Sawa, H., Inumaru, K.* High-Pressure Synthesis and Structural Characterization of the Type II Clathrate Compound $Na_{30.5}Si_{136}$ Encapsulating Two Sodium Atoms in the Same Silicon Polyhedral Cages. // J. Amer. Chem. Soc. - 2014. - **136**, N 21. - P. 7717-7725.

IMPMC, UPMC Sorbonne Universités, CNRS, MNHN, IRD, Paris, France
LACReSNE, Faculté des Sciences de Bizerte, Zarzouna, Tunisia
European Synchrotron Radiation Facility, Grenoble, France
Synchrotron SOLEIL, Gif-sur-Yvette, France